\documentclass[aps,showpacs,pra,superscriptaddress,]{revtex4}
\usepackage{amsmath}
\usepackage{amsfonts}
\usepackage{amssymb}
\usepackage{bm}

\usepackage{graphicx}

\begin{document}

\title{Propagation of dipole solitons in inhomogeneous highly dispersive
optical fiber media}
\author{ Houria Triki}
\affiliation{Radiation Physics Laboratory, Department of Physics, Faculty of Sciences,
Badji Mokhtar University, P. O. Box 12, 23000 Annaba, Algeria}
\author{Vladimir I. Kruglov}
\affiliation{Centre for Engineering Quantum Systems, School of Mathematics and Physics,
The University of Queensland, Brisbane, Queensland 4072, Australia}

\begin{abstract}
We consider the ultrashort light pulse propagation through an inhomogeneous
monomodal optical fiber exhibiting higher-order dispersive effects. Wave
propagation is governed by a generalized nonlinear Schr\"{o}dinger equation
with varying second-, third-, and fourth-order dispersions, cubic
nonlinearity, and linear gain or loss. We construct a new type of exact
self-similar soliton solutions that takes the structure of dipole via a
similarity transformation connected to the related constant-coefficients
one. The conditions on the optical fiber parameters for the existence of
these self-similar structures are also given. The results show that the
contribution of all orders of dispersion is an important feature to form
this kind of self-similar dipole pulse shape. The dynamic behaviors of the
self-similar dipole solitons in a periodic distributed amplification system
are analyzed. The significance of the obtained self-similar pulses is also
discussed.
\end{abstract}

\maketitle
\affiliation{$^{1}${\small Radiation Physics Laboratory, Department of Physics, Faculty
of Sciences, Badji Mokhtar University, P. O. Box 12, 23000 Annaba, Algeria}\\
$^{2}${\small Centre for Engineering Quantum Systems, School of Mathematics
and Physics, The University of Queensland, B Evolution of the dipole
self-similar intensity wave profile $\left\vert U(s,t)\right\vert ^{2}$ as
computed from Eq. (44) when $\Gamma (s)=\sin (s)$. The other parameters are
the same as in Fig. 1 except $A_{0}=0.1$.risbane, Queensland 4072, Australia}}

\section{Introduction}

Light pulses in optical fibers are referred to generically as solitons and
are usually described within the framework of the cubic nonlinear Schr\"{o}%
dinger equation (NLSE) that includes only basic effects on waves such as
group velocity dispersion (GVD) and self-phase modulation (SPM) \cite%
{H1,H2,A}. Depending on the anomalous dispersion or the normal dispersion,
the NLSE model allows for either bright or dark solitons, respectively \cite%
{Ha}. The formation of these solitons result from the exact balance between
the GVD and SPM of the material. Due to the robust and stable nature of
solitons, such wave packets have been successfully used as the information
carriers (optical bits) to transmit digital signals over long propagation
distances. However, many applications in various areas such as
ultrahigh-bit-rate optical communication systems, infrared time-resolved
spectroscopy, ultrafast physical processes, and optical sampling systems
require ultrashort (femtosecond) pulses \cite{A,Alka}, that leads to the
appearance of different higher-order effects in the optical material. For
instance, the third-order dispersion plays a significant role in propagation
if short pulses whose widths are nearly 50 femtosecond have to be injected 
\cite{P1,P2}. The fourth-order dispersion is also important when pulses are
shorter than 10 femtosecond \cite{P1,P2}. In such a situation, the wave
dynamics can be described by the higher-order nonlinear Schr\"{o}dinger
equation incorporating the contribution of various physical phenomena on
short-pulse propagation and generation.\ One notes that higher-order effects
introduce several new important phenomena into the system dynamics that are
absent in nonlinear media of the Kerr type.

In any material, however, there exists always some nonuniformities. Among
the many factors that cause nonuniformities in optical fiber systems \cite%
{Wan,Li,Mah}, we note (i) the variation in the lattice parameters of the
fiber media, (ii) the imperfection of manufacture, and (iii) the fluctuation
of the fiber diameters. These nonuniformities may significantly influence
various effects in the nonlinear medium such as dispersion, phase
modulation, gain (or loss), and others \cite{Li}. From the theoretical
standpoint, the description of the optical pulse propagation in
inhomogeneous fibers is generally based on the generalized NLSE with
distributed coefficients [i.e., allowing nonlinearity, dispersion, and gain
profiles to change with the distance along the direction of propagation] 
\cite{Kruglov,Kruglov1}. Such generalized model possesses a rich variety of
exact self-similar solutions that are characterized by a linear chirp \cite%
{Kruglov,Kruglov1,Serkin}. These self-similar pulses (also called
`similaritons') have attracted considerable interest in recent years because
of their extensive applications in photonics and fiber-optic
telecommunications \cite{Dumitru}. Importantly, these similaritons can
maintain the overall shape while allowing their parameters such as
amplitudes and widths to change with the modulation of system parameters 
\cite{Porsezian}.

Recently, propagation of self-similar waves has drawn much interest and many
important results have been presented, which is an essential prerequisite
for understanding the dynamical processes and mechanism of the complicated
phenomena in different inhomogeneous media \cite{Dai1}-\cite{Serkin1}. For
example,{\large \ }Dai et al. \cite{Dai1} investigated the dynamic behaviors
of spatial similaritons in inhomogeneous nonlinear cubic-quintic media. They
also discussed controllable optical rogue waves in the femtosecond regime 
\cite{Dai2}.{\large \ }In Ref. \cite{Dai3}, the authors constructed explicit
chirped and chirp-free self-similar solitary wave and cnoidal wave solutions
of the generalized cubic-quintic NLSE by applying the similarity
transformation method. Very recently, Pal et al. \cite{Pal} found
self-similar wave solutions of the quadratic-cubic NLSE describing wave
propagation through a tapered graded-index waveguide. Choudhuri et al. \cite%
{Triki1} derived the exact self-similar localized pulse solutions for the
NLSE with distributed cubic-quintic nonlinearities.\ Triki et al.\ \cite%
{Triki2} investigated the propagation of self-similar optical solitons on a
continuous-wave background in a quadratic-cubic non-centrosymmetric
waveguide. Liu et al. \cite{Liu} constructed a variety of spatiotemporal
self-similar wave solutions for the (3 +1)-dimensional variable-coefficients
NLSE with cubic-quintic nonlinearities. Serkin et al. \cite{Serkin}
discovered solitary nonlinear Bloch waves of the bright and dark types in
dispersion managed fiber systems and soliton lasers.

However, most investigations on the propagation of self-similar solitons in
inhomogeneous fiber systems have been focused on bright, dark, and kink type
self-similar solitary waves or solitons, as well as rogue and cnoidal waves.
But many novel localized structures including for example dipole solitons,
vortex solitons, and soliton trains have been demonstrated experimentally
and theoretically in both one- and two-dimensional nonlinear media \cite{JK}%
- \cite{XW}. To our knowledge, no exact self-similar \textit{`dipole'}
soliton solutions have been previously reported within the framework of the
variable-coefficient NLSE models. Moreover, the control of self-similar
localized pulses under the combined influence of GVD, TOD, FOD, and Kerr
nonlinearity management has not been widespread. In this paper, we
demonstrate the existence of self-similar pulses that takes a dipole
structure in inhomogeneous highly dispersive optical fibers and investigate
their propagation dynamics for different parameters.

Our results are presented as follows. Section II presents the method used
for obtaining traveling wave solutions of the extended NLSE that describes
the propagation of extremely short pulses inside a highly dispersive optical
fiber medium. In Sec. III, we derive analytical bright and dipole soliton
solutions of the model and their characteristics. In Sec. IV, the variation
of fiber dispersions, nonlinearity, and gain or loss is considered and a
similarity transformation to reduce the generalized nonlinear Schr\"{o}%
dinger equation with varying coefficients to the related
constant-coefficients one is presented. Self-similar dipole structures of
the generalized NLSE and their dynamical behaviors in a periodic distributed
amplification system are reported in Sec. V. Conclusions and future research
directions are addressed in Sec. VI.

\section{Traveling waves}

In this section, we reduce the
extended NLSE to an ordinary differential equation. The extended NLSE is
derived for the assumptions of slowly varying envelope, instantaneous
nonlinear response, and no higher order nonlinearities \cite{Cavalcanti}.
This nonlinear equation has the next form for the optical pulse envelope $%
E(z,\tau )$, 
\begin{equation}
iE_{z}=\alpha E_{\tau \tau }+i\sigma E_{\tau \tau \tau }-\epsilon E_{\tau
\tau \tau \tau }-\gamma \left\vert E\right\vert ^{2}E,  \label{1}
\end{equation}%
where $z$ is the longitudinal coordinate, $\tau =t-\beta _{1}z$ is the
retarded time, and $\alpha =\beta _{2}/2$, $\sigma =\beta _{3}/6$, $\epsilon
=\beta _{4}/24$, and $\gamma $ is the nonlinear parameter. The parameters $%
\beta _{k}=(d^{k}\beta /d\omega ^{k})_{\omega =\omega _{0}}$ are the k-order
dispersion of the optical fiber and $\beta (\omega )$ is the propagation
constant depending on the optical frequency.

This equation has been intensively studied for its importance from various
view points \cite{Cavalcanti}-\cite{Shagalov}. In particular, the
modulational instability phenomena of Eq. (\ref{1}) have been analyzed in
the region of the minimum group-velocity dispersion in \cite{Cavalcanti}. In
a very recent work, Kruglov and Harvey \cite{Kruglov2} presented an exact
solitary wave solution having the functional form of  `${sech}^{2}$'
for the NLSE (\ref{1}) including second-, third-, and fourth-order
dispersion effects. Moreover, Karpman et al. \cite{K1,K2} have studied the
time behavior of the amplitudes, velocities, and other parameters of
radiating solitons. Shagalov \cite{Shagalov} has investigated the effect of
the third- and fourth-order dispersions on the modulational instability. Roy
et al. \cite{Roy} have also studied the role of TOD and FOD in the radiation
emitted by fundamental soliton pulses in the form of dispersive waves within
the framework of the dimensionless form of the NLSE (\ref{1}).

We consider the solution of the generalized NLSE in the form, 
\begin{equation}
E(z,\tau )=u(x)\exp [i(\kappa z-\delta \tau +\theta )],  \label{2}
\end{equation}%
where $u(x)$ is a real function depending on the variable $x=\tau -qz$, and $%
q=v^{-1}$ is the inverse velocity. Also, $\kappa $ and $\delta $ are the
respective real parameters describing the wave number and frequency shift,
while $\theta $ represents the phase of the pulse at $z=0$.

Equations (\ref{1}) and (\ref{2}) lead to the next system of the ordinary
differential equations, 
\begin{equation}
(\sigma +4\epsilon \delta )\frac{d^{3}u}{d x^{3}}+(q-2\alpha \delta -3\sigma
\delta ^{2}-4\epsilon \delta ^{3})\frac{du}{dx}=0,  \label{3}
\end{equation}%
\begin{equation}
\epsilon \frac{d^{4}u}{dx^{4}}-(\alpha +3\sigma \delta +6\epsilon \delta
^{2})\frac{d^{2}u}{dx^{2}}+\gamma u^{3}-(\kappa -\alpha \delta ^{2}-\sigma
\delta ^{3}-\epsilon \delta ^{4})u=0.~~~~~~~~~~~~~~~~  \label{4}
\end{equation}%
In the general case the system of Eqs. (\ref{3}) and (\ref{4}) is
overdetermined because we have two differential equations for the function $%
u(x)$. However, if some constraints for the parameters in Eq. (\ref{3}) are
fulfilled the system of Eqs. (\ref{3}) and (\ref{4}) has non-trivial
solutions. We refer the solution of the extended NLSE where the function $%
E(z,\tau )$ is given by Eq. (\ref{2}) with $u(x) \neq \mathrm{constant}$ as
non-plain wave or traveling wave solution.

The system of Eqs. (\ref{3}) and (\ref{4}) with $\epsilon \neq 0$ yields the
non-plain wave solutions if and only if the next relations are satisfied: 
\begin{equation}
q=2\alpha \delta +3\sigma \delta ^{2}+4\epsilon \delta ^{3},~~~~\delta =-%
\frac{\sigma }{4\epsilon }.  \label{5}
\end{equation}%
The system of Eqs. (\ref{3}) and (\ref{4}) with $\epsilon =0$ has the
non-plain wave solutions only when the parameter $\sigma =0$. Note that Eq. (%
\ref{3}) is satisfied for an arbitrary function $\mathrm{u}(\mathrm{x})$
according to the conditions in Eq. (\ref{5}) with $\epsilon \neq 0$. The
relations in Eq. (\ref{5}) lead to the next expression for the velocity $%
v=1/q$ defined in the retarded frame, 
\begin{equation}
v=\frac{8\epsilon ^{2}}{\sigma (\sigma ^{2}-4\alpha \epsilon )}.  \label{6}
\end{equation}%
The relations in Eq. (\ref{5}) reduce the system of Eqs. (\ref{3}) and (\ref%
{4}) to the ordinary nonlinear differential equation, 
\begin{equation}
\epsilon \frac{d^{4}u}{dx^{4}}+b\frac{d^{2}u}{dx^{2}}-cu+\gamma
u^{3}=0,~~~~~~~  \label{7}
\end{equation}%
where the parameters $b$ and $c$ are 
\begin{equation}
b=\frac{3\sigma ^{2}}{8\epsilon }-\alpha ,~~~~c=\kappa +\frac{\sigma ^{2}}{%
16\epsilon ^{2}}\left( \frac{3\sigma ^{2}}{16\epsilon }-\alpha \right) .
\label{8}
\end{equation}

By analytically solving Eq. (\ref{7}), we obtain the soliton structures that
can propagate in the highly dispersive fiber medium. However, it would be
very difficult to find the closed form solutions of such ordinary
differential equation in which two even-order derivative terms coexist.
Obtaining solutions in analytic form is of great interest since these are
useful for instance to compare experimental results with theory. In
following, families of soliton solutions having the functional form of `$%
\mathrm{sech}^{2}(.)$' and `$\mathrm{sech}(.)\mathrm{th}(.)$' are derived in
presence of all physical parameters.

\section{Dipole soliton in highly dispersive optical fiber}

In this section, we consider the traveling wave solutions of Eq. (\ref{7})
in the form,

\begin{equation}
u(x)=\frac{F_{N}(x)}{G_{M}(x)}=\frac{\sum_{n=-N}^{N} A_{n}\exp[-nw(x-\eta)]}{%
\sum_{n=-M}^{M} B_{n}\exp[-nw(x-\eta)]}.  \label{9}
\end{equation}
The quartic dark soliton solution of this form is given by 
\begin{equation}
u(x)=A+B~\mathrm{th}^{2}[w(x-\eta )]=D-\frac{B}{\mathrm{ch}^{2}[w(x-\eta )]},
\label{10}
\end{equation}%
where $D=A+B$ and $D\neq 0$. In the case when $D=0$ we have the $sech^2$
solitary wave.

Substituting the function (\ref{10}) into Eq. (\ref{7}) and setting the
coefficients of independent terms equal to zero, we obtain the following
equations: 
\begin{equation}
cD=\gamma D^{3},~~~~c=16\epsilon w^{4}+4bw^{2}+3\gamma D^{2},  \label{11}
\end{equation}%
\begin{equation}
40\epsilon w^{4}+2bw^{2}+\gamma DB=0,~~~~120\epsilon w^{4}+\gamma B^{2}=0.
\label{12}
\end{equation}%
We now discuss solutions to these parametric equations for two cases: (1) for $D=0$, and  (2) for $D\neq 0$.
In the case (1) with $D=0$ and $E_{0}=-B$ we have 
\begin{equation}
w=\frac{1}{4}\sqrt{\frac{8\alpha \epsilon -3\sigma ^{2}}{10\epsilon ^{2}}}%
,~~~~E_{0}=\pm \sqrt{\frac{-3}{10\gamma \epsilon }}\left( \frac{3\sigma ^{2}%
}{8\epsilon }-\alpha \right) .  \label{13}
\end{equation}%
Further, we get from Eqs. (\ref{11}) and (\ref{12}) a
condition on the parameter $c$ as 
\begin{equation}
c=-\frac{4}{25\epsilon }\left( \frac{3\sigma ^{2}}{8\epsilon }-\alpha
\right) ^{2}.  \label{14}
\end{equation}

Incorporating these results into Eq. (\ref{10}), we obtain the following
soliton solution to the extended NLSE (\ref{1}) \cite{Kruglov2}:

\begin{equation}
E(z,\tau )=E_{0}~\mathrm{sech}^{2}(w\xi )\exp [i(\kappa z-\delta \tau
+\theta )],  \label{15}
\end{equation}%
where $\xi =\tau -v^{-1}z-\eta $, with $\eta $ being the position of the
pulse at $z=0$. The wave number $\kappa $ follows from Eqs. (\ref{8}) and (%
\ref{14}) as 
\begin{equation}
\kappa =-\frac{4}{25\epsilon ^{3}}\left( \frac{3\sigma ^{2}}{8}-\alpha
\epsilon \right) ^{2}-\frac{\sigma ^{2}}{16\epsilon ^{3}}\left( \frac{%
3\sigma ^{2}}{16}-\alpha \epsilon \right) .  \label{16}
\end{equation}
The velocity $v$ and frequency shift $\delta$ in this soliton solution are
given by Eqs. (\ref{5}) and (\ref{6}).

Physically, Eq. (\ref{15}) describes a bright stationary pulse with
amplitude $E_{0}$ and inverse temporal width $w$\ depending on all order of
dispersion as well as nonlinearity. It follows from Eq. (\ref{13}) that this 
$\mathrm{sech}^{2}$ solitary wave exists when the next two conditions are
satisfied: $\gamma \epsilon <0$, $8\alpha \epsilon >3\sigma ^{2}$.

In the case (2) with $D\neq 0,$ Eqs. (\ref{11}) and (\ref{12}) lead to the
complex values for parameters $B$ and $D$. However the function $u(x)$ in
Eq. (\ref{7}) is real which contradicts to such complex parameters. Thus the
extended NLSE given by Eq. (\ref{1}) does not have a quartic dark soliton
solution.

We have also found that Eq. (\ref{7}) admits an exact dipole soliton
solution of the form, 
\begin{equation}
u(x)=E_{0}\frac{\mathrm{sh}[w(x-\eta )]}{\mathrm{ch}^{2}[w(x-\eta )]}.
\label{17}
\end{equation}%
Inserting this solution into Eq. (\ref{7}) and equating the coefficients of
independent terms, one obtains 
\begin{equation}
c=\epsilon w^{4}+bw^{2},~~~~120\epsilon w^{4}=\gamma E_{0}^{2},  \label{18}
\end{equation}%
\begin{equation}
60\epsilon w^{4}+6bw^{2}-\gamma E_{0}^{2}=0.  \label{19}
\end{equation}%
These equations yield the dipole soliton solution as 
\begin{equation}
E(z,\tau )=E_{0}~\mathrm{sech}(w\xi )~\mathrm{th}(w\xi )\exp [i(\kappa
z-\delta \tau +\theta )],  \label{20}
\end{equation}%
where $\xi =\tau -v^{-1}z-\eta $. Thus we have the following relations for
the pulse inverse width $w$ and amplitude $E_{0}$, 
\begin{equation}
w=\frac{1}{4}\sqrt{\frac{3\sigma ^{2}-8\alpha \epsilon }{5\epsilon ^{2}}}%
,~~~~E_{0}=\pm \sqrt{\frac{6}{5\gamma \epsilon }}\left( \frac{3\sigma ^{2}}{%
8\epsilon }-\alpha \right) .  \label{21}
\end{equation}%
Eqs. (\ref{18}) and (\ref{19}) yield the parameter $c=11b^{2}/100\epsilon .$
Hence it follows from Eq. (\ref{8}) that the wave number $\kappa $ in the
dipole soliton solution is given by 
\begin{equation}
\kappa =\frac{11}{100\epsilon ^{3}}\left( \frac{3\sigma ^{2}}{8}-\alpha
\epsilon \right) ^{2}-\frac{\sigma ^{2}}{16\epsilon ^{3}}\left( \frac{%
3\sigma ^{2}}{16}-\alpha \epsilon \right) .  \label{22}
\end{equation}%
The velocity $v$ and frequency shift $\delta$ in this dipole soliton
solution are given by Eqs. (\ref{5}) and (\ref{6}). It follows from Eq. (\ref%
{21}) that the dipole soliton solution exists when the next two conditions
are satisfied: $\gamma \epsilon >0$, $3\sigma ^{2}>8\alpha \epsilon $.

The corresponding energy $\mathcal{E}$ of the dipole solitons is given by 
\begin{equation}
\mathcal{E}=\int_{-\infty }^{+\infty }|E(z,\tau )|^{2}d\tau =\frac{(3\sigma
^{2}-8\alpha \epsilon )^{3/2}}{4\gamma \epsilon \sqrt{5\epsilon ^{2}}}.
\label{23}
\end{equation}
Note that the energy of the pulse $\mathcal{E}$ is the integral of motion [$d%
\mathcal{E}/dz=0$] of the NLSE (\ref{1}) for any pulses satisfying the
boundary condition: $E(z,\tau )\rightarrow 0$ for $\tau\rightarrow \pm\infty$%
.

\section{Transformation to generalized NLSE with variable coefficients}

We consider in this section the variations of fiber dispersion,
nonlinearity, and gain or loss. For our purpose, the dynamics of pulses is
described by the following generalized NLSE with distributed coefficients: 
\begin{equation}
iU_{s}=D(s)U_{tt}+iP(s)U_{ttt}-Q(s)U_{tttt}-R(s)\left\vert U\right\vert
^{2}U+i\Gamma (s)U.~~~~~~~~~~~~~~~~~  \label{24}
\end{equation}%
where $D(s),$ $P(s)$ and $Q(s)$ are the variable GVD, TOD, and FOD
coefficients, respectively. The function $R(s)$ stands for the varying Kerr
nonlinearity coefficient, while $\Gamma (s)$ denotes the amplification $%
\left( \Gamma (s)>0\right) $ or absorption coefficient $\left( \Gamma
(s)<0\right) $.

In the simplest case, when all the coefficients are constants and $\Gamma
(s)=0$, then Eq. (\ref{24}) can be transformed into the constant-coefficient
NLSE (\ref{1}). It is of interest to control optical solitons in
communication systems when all orders of dispersion, nonlinearity, and gain
or loss are varied as described by the NLSE (\ref{24}). In following, we
first search for exact self-similar soliton solutions of the
variable-coefficient NLSE (\ref{24}) by employing the similarity
transformation method and then discuss their propagation behaviors in a
specified soliton control system.

\begin{figure}
\includegraphics[width=\linewidth,trim={3cm 15cm 3cm 7cm},clip]{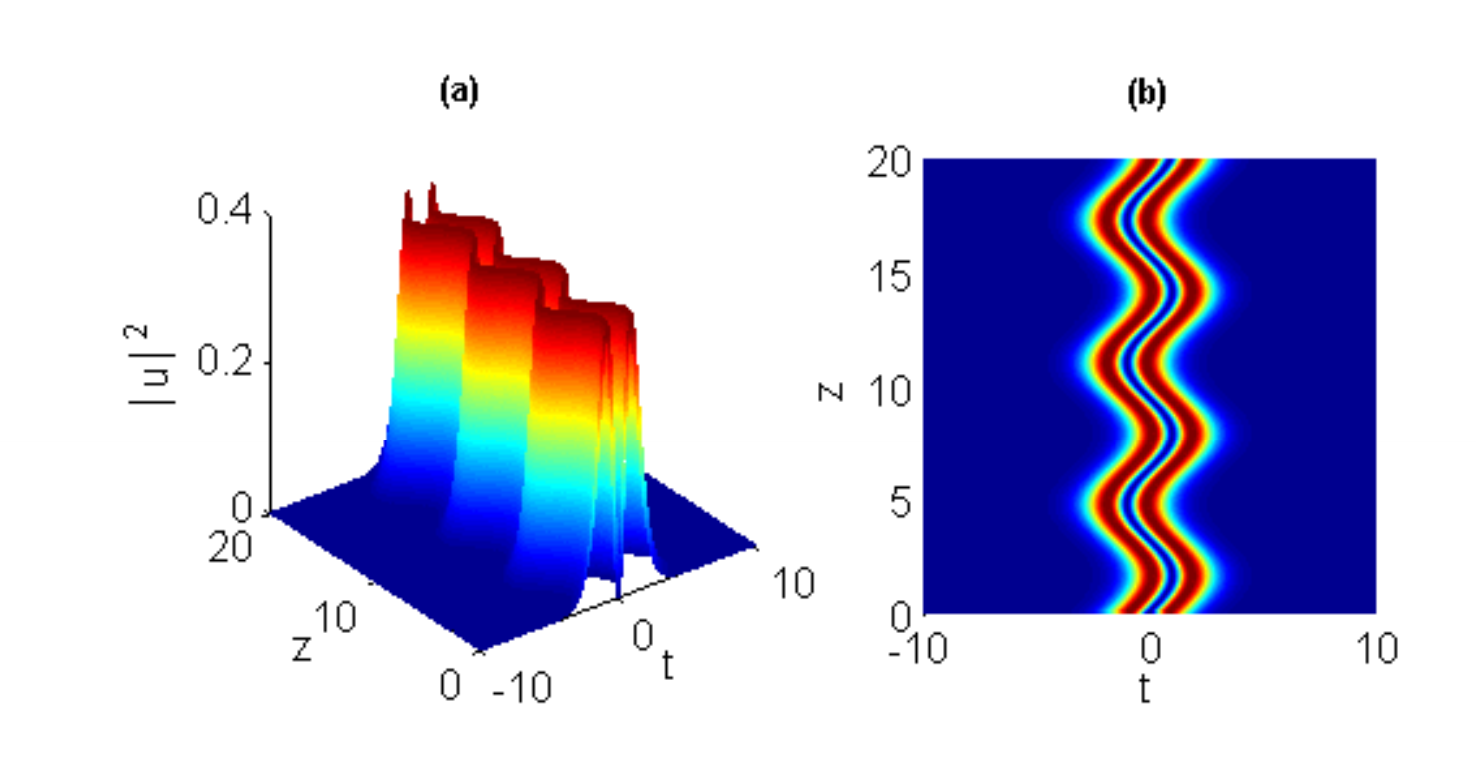}
\caption{Evolution of the dipole self-similar intensity wave profile 
$\left\vert U(s,t)\right\vert ^{2}$ as computed from Eq. (44). The
parameters are defined in the text.} 
\label{FIG.1.}
\end{figure}

\begin{figure}
\includegraphics[width=\linewidth,trim={3cm 15cm 3cm 7cm},clip]{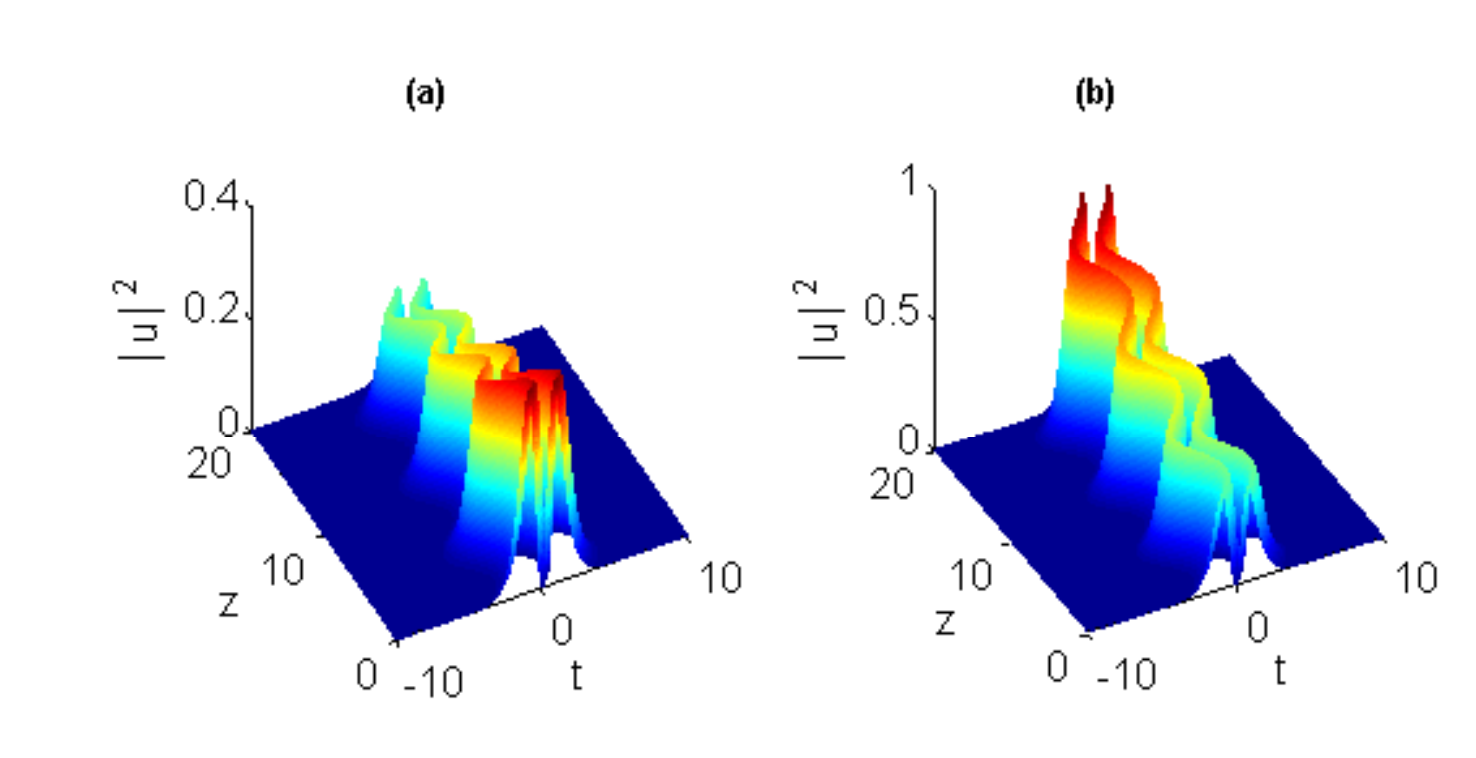}
\caption{Evolution of the dipole self-similar intensity wave profile 
$\left\vert U(s,t)\right\vert ^{2}$ as computed from Eq. (44) when (a) $%
\Gamma _{0}=-0.02$, (b) $\Gamma _{0}=0.02$. The other parameters are the
same as in Fig. 1.} 
\label{FIG.2.}
\end{figure}

\begin{figure}
\includegraphics[width=\linewidth,trim={3cm 15cm 3cm 7cm},clip]{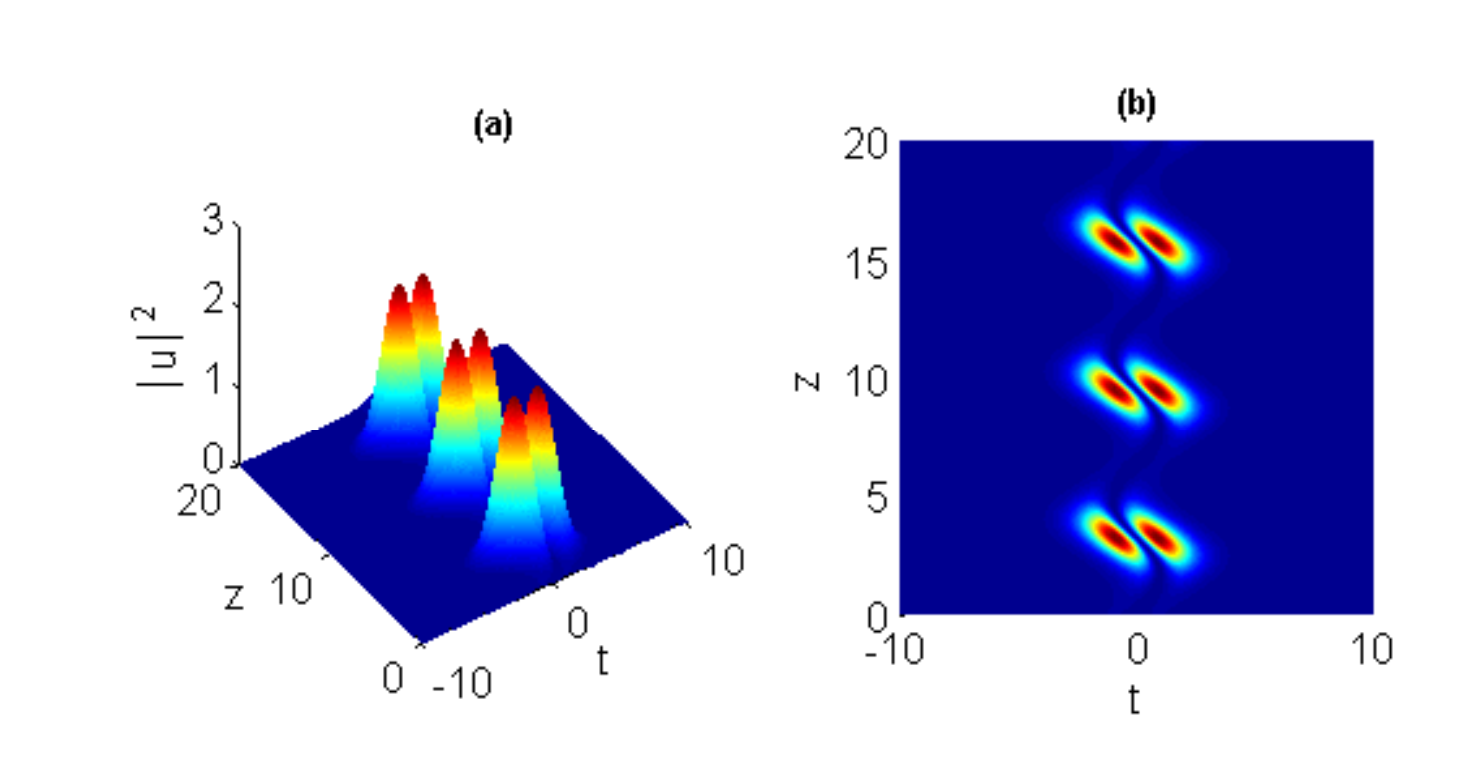}
\caption{ Evolution of the dipole self-similar intensity wave profile 
$\left\vert U(s,t)\right\vert ^{2}$ as computed from Eq. (44) when $\Gamma
(s)=\sin (s)$. The other parameters are the same as in Fig. 1 except $%
A_{0}=0.1$.} 
\label{FIG.3.}
\end{figure}

\begin{figure}
\includegraphics[width=\linewidth,trim={3cm 15cm 3cm 7cm},clip]{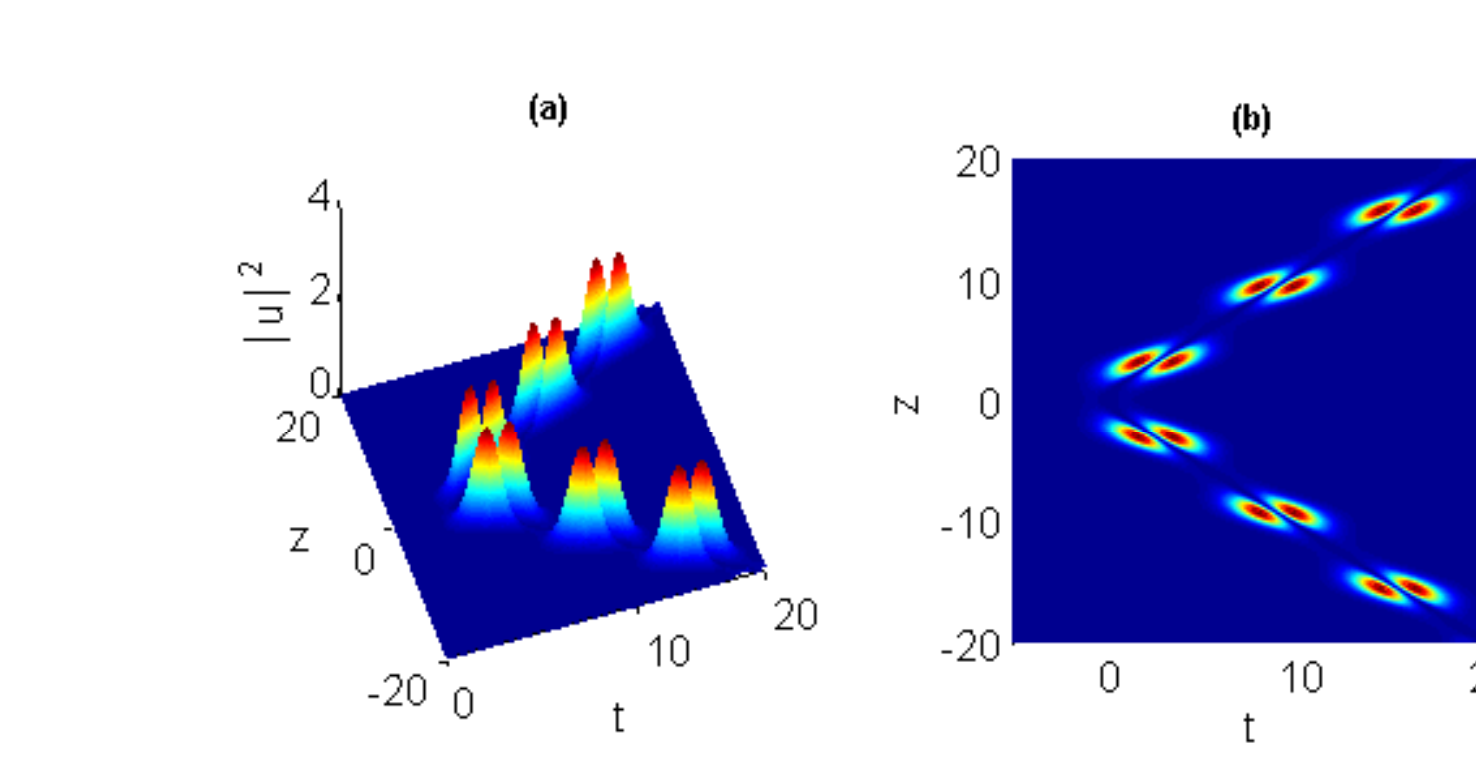}
\caption{Evolution of the dipole self-similar intensity wave profile 
$\left\vert U(s,t)\right\vert ^{2}$ as computed from Eq. (44) when $\Gamma
(s)=\sin (s)$ and $Q(s)=\tanh(s).$ The other parameters are the same as in
Fig. 1 except $A_{0}=0.1$.} 
\label{FIG.4.}
\end{figure}

We first construct the transformation \cite{Dai1}-\cite{Pal}%
\begin{equation}
U(s,t)=A(s)E[z(s),\tau (s,t)]e^{i\phi (s,t)},  \label{25}
\end{equation}%
where $A(s)$\ is the amplitude, $E(z,\tau )$ is the optical pulse envelope, $%
z=z(s)$ and $\tau =\tau (s,t)$ are two unknown functions to be determined,
and $\phi (s,t)$\ is the phase function.

Upon substituting Eq. (\ref{25}) into Eq. (\ref{24}) leads to (\ref{1}), but
now we must have the following set of equations:%
\begin{eqnarray}
&&\left. A_{s}-\Gamma A-DA\phi _{tt}+3PA\phi _{t}\phi _{tt}+QA\phi
_{tttt}-6QA\phi _{t}^{2}\phi _{tt}=0,\right.  \label{26} \\
&&\left. \tau _{s}-2D\tau _{t}\phi _{t}+3P\tau _{t}\phi _{t}^{2}-P\tau
_{ttt}+4Q\tau _{ttt}\phi _{t}+6Q\tau _{tt}\phi _{tt}-4Q\tau _{t}\phi
_{t}^{3}+4Q\tau _{t}\phi _{ttt}=0,\right.  \label{27} \\
&&\left. \phi _{s}-D\phi _{t}^{2}-P\phi _{ttt}+P\phi _{t}^{3}+4Q\phi
_{t}\phi _{ttt}-Q\phi _{t}^{4}+3Q\phi _{tt}^{2}=0,\right.  \label{28} \\
&&\left. \left( 3P\phi _{t}-D\right) \tau _{tt}+3P\tau _{t}\phi _{tt}+Q\tau
_{tttt}-12Q\tau _{t}\phi _{t}\phi _{tt}-6Q\tau _{tt}\phi _{t}^{2}=0,\right.
\label{29} \\
&&\left. -3P\tau _{t}\tau _{tt}+12Q\tau _{t}\tau _{tt}\phi _{t}+6Q\tau
_{t}^{2}\phi _{tt}=0,\right.  \label{30} \\
&&\left. RA^{2}=\gamma z_{s},\right.  \label{31} \\
&&\left. \left( D-3P\phi _{t}\right) \tau _{t}^{2}-4Q\tau _{t}\tau
_{ttt}+6Q\tau _{t}^{2}\phi _{t}^{2}-3Q\tau _{tt}^{2}=\alpha z_{s},\right.
\label{32} \\
&&\left. -P\tau _{t}^{3}+\sigma z_{s}+4Q\tau _{t}^{3}\phi _{t}=0,\right.
\label{33} \\
&&\left. Q\tau _{t}^{4}=\epsilon z_{s},\right.  \label{34} \\
&&\left. \tau _{tt}=0,\right.  \label{35}
\end{eqnarray}

\noindent Solving these equations self-consistently allows us to find the
following parameters that characterize the self-similar pulse:%
\begin{eqnarray}
&&\left. A(s)=A_{0}\exp \left[ \int_{0}^{s}\Gamma (\zeta )d\zeta \right]
,\right.  \label{36} \\
&&\left. \tau (s,t)=k\left[ t+p\left( 4p^{2}+\frac{2\alpha k^{2}+3p\sigma k}{%
\epsilon }\right) \int_{0}^{s}Q(\zeta )d\zeta \right] +t_{0},\right.
\label{37} \\
&&\left. z(s)=\frac{k^{4}}{\epsilon }\int_{0}^{s}Q(\zeta )d\zeta ,\right.
\label{38} \\
&&\left. \phi (s,t)=p\left[ t+p\left( 3p^{2}+\frac{\alpha k^{2}+2p\sigma k)}{%
\epsilon }\right) \int_{0}^{s}Q(\zeta )d\zeta \right] +\phi _{0},\right.
\label{39}
\end{eqnarray}

\noindent where $k$ and $p$ are parameters relative to pulse width and phase
shift, respectively, $\tau (s,t)$ is the mapping variable, and $z(s)$
represents the effective propagation distance. Here the subscript $0$
denotes the initial values of the corresponding parameters at distance $s=0.$
Furthermore, the constraint conditions on the inhomogeneous fiber parameters
are given as%
\begin{eqnarray}
&&\left. D(s)=\left( 6p^{2}+\frac{\alpha k^{2}+3p\sigma k}{\epsilon }\right)
Q(s),\right.  \label{40} \\
&&\left. R(s)=\frac{\gamma k^{4}}{\epsilon A_{0}^{2}}\exp \left[
-2\int_{0}^{s}\Gamma (\zeta )d\zeta \right] Q(s),\right.  \label{41} \\
&&\left. P(s)=\left( 4p+\frac{\sigma k}{\epsilon }\right) Q(s).\right.
\label{42}
\end{eqnarray}

Equations (\ref{38}) and (\ref{39}) show that the effective propagation
distance and phase are strongly dependent on the FOD parameter $Q(s)$. The
latter influences the GVD coefficient $D(s)$, Kerr nonlinear coefficient $%
R(s),$ and TOD coefficient $P(s)$ as seen from the preceding constraints.
Hence, one can control the dynamics of propagating self-similar dipole
solitons in the fiber medium by selecting the profile of this parameter
suitably.

Thus, the general form of self-similar solutions of the generalized NLSE (%
\ref{24}) is of the form%
\begin{eqnarray}
U(s,t) &=&A_{0}E\left\{ \frac{k^{4}}{\epsilon }\int_{0}^{s}Q(\zeta )d\zeta ,k%
\left[ t+p\left( 4p^{2}+\frac{2\alpha k^{2}+3p\sigma k}{\epsilon }\right)
\int_{0}^{s}Q(\zeta )d\zeta \right] +t_{0}\right\}  \notag \\
&&\times \exp \left[ \int_{0}^{s}\Gamma (\zeta )d\zeta +i\phi (s,t)\right] .
\label{43}
\end{eqnarray}%
where the phase function $\phi (s,t)$ is given by Eq. (\ref{39}) and $%
E(z,\tau )$ are the exact solutions of Eq. (\ref{1}).

Therefore exact self-similar solutions to Eq. (\ref{24}) can be constructed
by using the exact solutions of Eq. (\ref{1}) via the transformation (\ref%
{43}). One thus needs to use the closed form solutions of the
constant-coefficient NLSE (\ref{1}) presented above.

\section{Self-similar dipole soliton solutions}

Making use of the exact solution given in Eq. (\ref{20}) of the extended
constant-coefficient NLSE (\ref{1}), the transformations in Eq. (\ref{43}),
and Eqs. (\ref{36})-(\ref{39}), we can construct the self-similar solutions
of the generalized NLSE with varying coefficients (\ref{24}).
The self-similar dipole soliton solution of Eq. (\ref{24}) is then given by%
\begin{equation}
U(s,t)=A_{0}E_{0}\exp \left[ \int_{0}^{s}\Gamma (\zeta )d\zeta \right] 
\mathrm{sech}(w\xi )~\mathrm{th}(w\xi )\exp \left[ i\Phi (s,t)\right] ,
\label{44}
\end{equation}%
where the traveling coordinate $\xi $ is given by%
\begin{equation}
\xi (s,t)=kt-\eta +\left\{ kp\left( 4p^{2}+\frac{2\alpha k^{2}+3p\sigma k}{%
\epsilon }\right) -\frac{k^{4}}{v\epsilon }\right\} \int_{0}^{s}Q(\zeta
)d\zeta +t_{0},  \label{45}
\end{equation}

\noindent and the phase of the field $\Phi $ has the form%
\begin{equation}
\Phi (z,t)=\kappa z-\delta \tau +\theta +\phi (s,t),  \label{46}
\end{equation}

\noindent where $\tau $ and $z$ are given by Eqs. (\ref{37}) and (\ref{38})
respectively, while the phase $\phi(s,t)$ is given by Eq. (\ref{39}).

From the results obtained above, we see that the contribution of all orders
of dispersion is necessary for the existence of self-similar dipole soliton
solutions for the generalized NLSE with distributed coefficients (\ref{24}).
This is markedly different from the dipole structures of many constant
coefficient NLSE models describing femtosecond pulse dynamics in homogeneous
fibers \cite{Amit}-\cite{Triki4}, which exist only when both GVD, TOD and
FOD are compensated. It should be noted that the simultaneous compensation
of various order of dispersion is generally more difficult in optical
systems \cite{Amit}. Therefore, our results could be of importance in
applications of dipole type solitons in optical fiber systems exhibiting
dispersive effects up to the fourth order.

To examine the dynamical evolution of the obtained self-similar structure in
the optical fiber medium, it is worthwhile to consider a specific soliton
control system. Here, we focus on studying the propagation of self-similar
waves through a periodically distributed amplification system similar to
that of \ Ref. \cite{Dai2}. Especially, we suppose that the FOD management
takes the form of a cosinelike space-dependent rapidly varying function as 
\cite{Dai2}: $Q(s)=d_{4}\cos (gz)$, while the gain function is given by $%
\Gamma (s)=\Gamma _{0}$. Here $d_{4}$ and $g$ are the parameters to describe
FOD and $\Gamma _{0}$ represents the constant net gain or loss. From the
practical point of view, the propagation with periodic dispersion is of
great importance as it has application in enhancing the signal to noise
ratio and reducing Gordon-Hauss time jitter and is also helpful in
suppressing the phase matched condition for four-wave mixing \cite%
{Konar,Loomba}. Then according to Eq. (\ref{38}), the effective propagation
distance can be obtained as $z(s)=\frac{d_{4}k^{4}}{\epsilon g}\sin (gz)$,
implying that $z$ varies periodically with\ the propagation distance $s$.
Furthermore, the amplitude can be calculated from Eq. (\ref{36}) as: $%
A(s)=A_{0}\exp \left( \Gamma _{0}s\right) .$ This means that the amplitude
of the self-similar pulse will undergo increase ($\Gamma _{0}>0$) and
decrease ( $\Gamma _{0}<0$) along the propagation distance, while it remains
a constant when the gain (loss) vanishes ($\Gamma _{0}=0$). As concerns the
other parameters, they can be obtained exactly through Eqs. (\ref{40}), (\ref%
{41}) and (\ref{42}).

Consider first the most interesting case when the optical fiber medium does
not subject to the effect of the gain or loss effect [i.e., $\Gamma _{0}=0$]$%
.$\ The evolution of the self-similar dipole soliton solution (\ref{44})
calculated with the framework of the generalized NLSE (\ref{24}) is shown in
Figs. 1(a) and 1(b) with the parametric values: $\alpha =-1,$\ $\gamma =2,$ $%
\sigma =1,$ and $\epsilon =\frac{1}{4}.$\ Also we take $k=p=1$, $\
A_{0}=0.3098,$\ $d_{4}=\frac{1}{8},\ \eta =0,\ g=1,$\ and $t_{0}=0$.\ From
these figures, one can clearly see that the self-similar structure display a
snakelike behavior along the propagation distance due to the presence of
periodic distributed dispersion parameter\ $Q(s).$\ For such oscillatory
trajectory, the self-similar pulse keep no change in propagating along
optical medium although its position oscillate periodically (which is called
\textquotedblleft Snakelike\textquotedblright\ in Ref. \cite{ZY}).

When the self-similar dipole soliton is subjected to the action of a
constant gain or loss, that is, $\Gamma (s)=\Gamma _{0},$\ its intensity
decreases when $\Gamma _{0}<0$\ and increase when $\Gamma _{0}>0$, and the
time shift and the group velocity of the soliton pulse are changing while
the soliton keeps its shape in propagation along the fiber [Figs. 2(a)-(b)].
One readily concludes that the gain parameter affects only the evolution of
soliton peak and has no influence on the width or shape of the pulse.

Another interesting behavior appears when the gain or loss function is
chosen to vary periodically with the propagation distance as $\Gamma
(s)=\sin (s).$\ This spatial profile of gain (loss) was first used in
studying soliton management in inhomogeneous pure Kerr media \cite{JF}. The
corresponding intensity profiles of self-similar dipole soliton are shown in
Figs. 3(a)-(b) for the same values of parameters as those in Fig.\ 1 except $%
A_{0}=0.1$. As can be seen from this figure, in the presence of periodic
gain, the dipole solitons emerge periodically in the inhomogeneous fiber
system.

Let us now investigate the propagation dynamics of self-similar dipole
pulses in a distributed fiber system whose FOD and gain or loss parameters
are distributed according to \cite{JF}: $Q(s)=\tanh (s)$\ and $\Gamma
(s)=\sin (s).$\ Figures 4(a)-(b) show the nonlinear evolution of the
self-similar solution (\ref{44}) for the same values of parameters as those
in Fig. 1 except $A_{0}=0.1$. We observe an interesting periodic occurrence
of dipole solitons appearing for this choice of dispersion and gain or loss
management, as can be seen from Fig. 4.

\section{Conclusion}
In this paper, we have investigated the variable-coefficient nonlinear Schr%
\"{o}dinger equation incorporating, at the highest order, a fourth-order
dispersion, which governs the femtosecond optical pulse propagation in an
inhomogeneous highly dispersive fiber media. We have first constructed the
relation between this generalized wave equation and the related
constant-coefficients one via a similarity transformation. Then based on the
obtained transformation, we have derived the exact self-similar dipole
soliton solutions of the considered model. Conditions on the varying optical
fiber parameters for the existence of these self-similar structures are also
presented. It is found that the existence of these self-similar dipole
solitons in an inhomogeneous highly dispersive\ optical fiber media
crucially depends, indeed, on all orders of dispersion. We have further
discussed the propagation dynamics of self-similar waves in dispersion
changing periodically fiber system. It is observed that the self-similar
wave structure and dynamical behavior can be controlled by choosing
appropriate parameters of fourth-order dispersion and gain or loss.

An issue of prime importance is the stability of self-similar solitons with
respect to perturbations. This is because only stable (or weakly unstable)
solitary waves are promising for experimental observations and practical
applications \cite{Triki1,Shi}. It should be noted that the stability
analysis can be achieved by numerical simulations and the linear stability
theory of the solutions with perturbations initially implanted. For the
present generalized nonlinear Schr\"{o}dinger equation model with varying
coefficients, we have found that the obtained self-similar dipole structures
essentially exist due to a balance among all order of dispersion and Kerr
nonlinearity effect. The stability aspects of such privileged localized
light pulses typically require detailed individual analysis based on such
balance aspects. Detailed stability analysis are now under investigation.



\end{document}